\documentclass[prb,nopacs,twocolumn,preprintnumbers,amsmath,amssymb]{revtex4-1}

\usepackage{epsfig}
\usepackage{graphicx}% Include figure files
\usepackage{dcolumn}% Align table columns on decimal point
\usepackage{color}
\usepackage{bm}% bold math

\begin{document}

\title{Interactions and magnetic moments near vacancies and resonant impurities in graphene}

\author{P. Haase$^1$, S. Fuchs$^1$, T. Pruschke$^1$, H. Ochoa$^2$, F. Guinea$^2$ }

{\affiliation{$^1$ Department of Physics, University of G\"ottingen, 37077 G\"ottingen, Germany. \\
$^2$ Instituto de Ciencia de Materiales de Madrid. CSIC. Sor Juana In\'es de la Cruz 3. 28049 Madrid. Spain.}

\begin{abstract}
The effect of electronic interactions in graphene with vacancies or resonant scatterers is investigated. We apply dynamical mean-field theory in combination with quantum Monte Carlo simulations, which allow us to treat non-perturbatively quantum fluctuations beyond Hartree-Fock approximations. The interactions narrow the width of the resonance and induce a Curie magnetic susceptibility, signaling the formation of local moments. The absence of saturation of the susceptibility at low temperatures suggests that the coupling between the local moment and the conduction electrons is ferromagnetic.
\end{abstract}
%\pacs{73.20.-r; 73.20.Hb; 73.23.-b; 73.43.-f}

\maketitle
{\em Introduction.}
Since its isolation\cite{Netal04,Netal05}, single layer graphene has attracted a great deal of attention, due to novel features. The massless nature of the charge carriers implies that the density of states vanishes at the Fermi energy in a neutral layer\cite{NGPNG09}. The low density of states in its vicinity allows for the formation of sharp resonances due to vacancies\cite{PGLPN06} or impurities like hydrogen which form a strong covalent bond with the carbon atoms, the so called resonant impurities\cite{WYLGK10}. These resonances have been observed in graphite surfaces\cite{UBGG10}.

The enhancement in the density of states by those resonances and the electron electron interaction favor the formation of local moments. The states associated to the resonances differ from those induced by coupled magnetic dopants in a number of ways: i) They are built up from the same $\pi$ orbitals as the conduction band of graphene, ii) The resonance state is orthogonal to the conduction states, and the hopping between the resonance and the extended states vanishes, and  iii) They extend over a large region near the defect, as there is no gap in the spectrum to confine them.

Mean field arguments based on the enhancement of the local density of states favor the formation of a static magnetic moment, as shown in a number of calculations\cite{DSL04,LFMKN04,KH07,PFB08,Y08,Y10}. It is known that quantum fluctuations, not included in Hartree-Fock approximations, determine the low temperature properties of magnetic impurities in solids, described by the ferro- or antiferromagnetic Kondo model\cite{H93}. The differences between the resonant levels in graphene and magnetic impurities in metals and graphene\cite{HG07,UKPN08,JK10} imply that  mean field results cannot be extrapolated in a straightforward way to low temperatures.

In the following, we analyze the electronic properties of graphene by non perturbative methods beyond a static mean field approximation. Our results show that, for reasonable interaction strengths, the main features are well described by assuming the existence of a fluctuating magnetic moment. This moment is not quenched at the lowest accessible temperatures, suggesting a ferromagnetic coupling with the conduction band. For a finite concentration of resonant impurities, the absence of competition between the antiferromagnetic Kondo effect and the RKKY interaction might lead to ferromagnetism, provided that the concentration of impurities is large enough\cite{Eetal03,Betal07,Oetal07,Setal10}.

\begin{figure}
\includegraphics[width=\columnwidth]{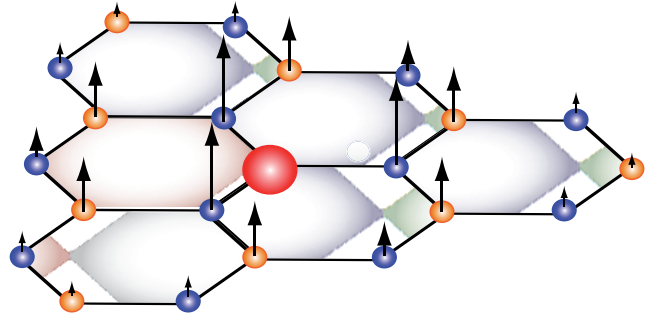}
\caption[fig]{(Color online). Sketch of the model studied in the text. The presence of a vacancy, or a resonant scatterer, passivates one of the lattice sites. The resulting localized state near the Dirac energy becomes spin polarized.}
 \label{sketch}
\end{figure}
{\em The model.}
We describe the $\pi$ band of graphene by a nearest neighbor tight binding model, with a hopping parameter $t$. The effect of a resonant scatterer strongly bound to a given site is taken into account by shifting the on-site energy by an amount $\epsilon_0$. In the limit $| \epsilon_0 | \gg t$ the model describes an unrelaxed vacancy. For $| \epsilon_0 | \gtrsim t$, a resonance near the Dirac energy builds up. This resonance moves to the Dirac energy and its width vanishes in the limit $| \epsilon_0 | / t \rightarrow \infty$. The main features of the model, including the possibility of a local moment at the resonance, are shown in Fig.~\ref{sketch}.

We assume that the long range part of the electron-electron interaction is screened, and we describe the electron electron interaction by an on-site Hubbard repulsive term, $U$. The full hamiltonian is
\begin{align}
  {\cal H} &= -t \sum_{\langle i, j \rangle \sigma} c^\dag_{i\sigma} c^{\phantom\dagger}_{j\sigma}
  - \epsilon_0 \sum_\sigma  c^\dag_{0\sigma} c^{\phantom\dagger}_{0\sigma} \nonumber \\
  &+ U \sum_i \left( n_{i\uparrow} - \frac{1}{2} \right) \left( n_{i\downarrow} - \frac{1}{2} \right) \;\;,
  \label{hamil}
\end{align}
where $c^{(\dagger)}_{i\sigma}$ annihilates (creates) an electron with spin $\sigma=\uparrow,\downarrow$ on lattice site $i$ and where $n_{i\sigma} = c^{\dagger}_{i\sigma} c^{\phantom\dagger}_{i\sigma}$ denotes the corresponding number density. The hopping $t$ is only finite for neighboring lattice sites (denoted by $\langle i, j\rangle$). The lattice site $i=0$ corresponds to the impurity where the local on-site energy $\epsilon_0$ is non-zero.
We study the model at half filling, i.\,e., at chemical potential $\mu=0$.

We approximate the interacting lattice problem by assuming that the impurity site where $\epsilon_0 \ne 0$ and a neighboring site are attached to an effective medium described by a local self-energy $\Sigma$ as sketched in Fig.~\ref{method}.
\begin{figure}
\includegraphics[width=\columnwidth]{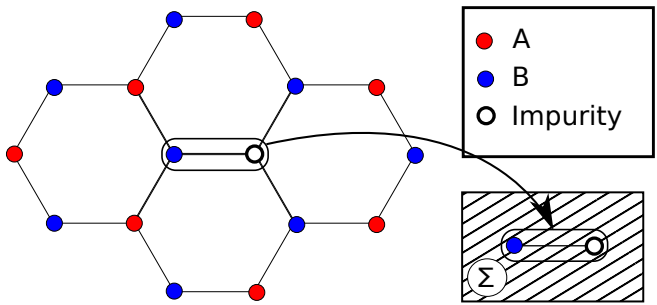}
\caption[fig]{(Color online). Sketch of the lattice system with two sub-lattices A and B (left panel). The impurity is denoted by an open circle. We map the lattice on a two-site cluster consisting of only one unit cell (right panel). The cluster is embedded in an interacting medium which determined by a local self-energy $\Sigma$. The self-energy is calculated either by DMFT or by second-order perturbation theory.}
 \label{method}
\end{figure}
The effect of the impurity on the bath vanishes in the thermodynamic limit and can therefore be neglected \cite{Hofstetter}.  The resulting two-site cluster problem is solved numerically by quantum Monte Carlo (QMC) simulations. We employ QMC methods in continuous imaginary time which perform a systematic expansion in the interaction term of the Hamiltonian \cite{rubtsov1, rubtsov2}. The QMC solution of the cluster problem is numerically exact and fully incorporates interactions and quantum fluctuations.  The analysis can be regarded as the solution of a quantum impurity problem where the interaction effects are included at the impurity site and at a close neighbor, and approximated by means of a local self-energy in the surrounding medium.

In order to calculate the self-energy necessary to determine the effective medium, we start by simulating the homogeneous lattice system with $\epsilon_0=0$. We calculate the self-energy necessary to obtain the effective medium using either Dynamical Mean Field Theory (DMFT) \cite{DMFT-bible} or second-order perturbation theory in $U/t$. DMFT fully incorporates quantum fluctuations local to the cluster but ignores spatial fluctuations. The medium depends on the self-energy of the cluster system and has to be calculated self-consistently by an iterative procedure using QMC simulations. Subsequently, the impurity is added to the system and one additional QMC simulation is performed using the medium of the converged DMFT calculation.

In order to check the quality of the DMFT approximation, we additionally calculate the self-energy of the homogeneous lattice system using second-order perturbation theory in $U/t$. This self-energy -- instead of the self-consistently determined DMFT solution -- is used to calculate the bath of the impurity problem which is then again solved by QMC. The perturbative self-energy does not include quantum fluctuations to all orders. However, it incorporates non-local effects of the actual lattice structure which are neglected by DMFT. Thus we are able to test the influence of non-local correlations and the accuracy of the DMFT approximation.

QMC methods map quantum-mechanical systems on a classical one at the expense of an additional dimension which -- in most cases -- is an imaginary time dimension. Thus, QMC can only provide dynamical data for imaginary times or frequencies. The necessary analytic continuation to physically relevant real times or frequencies is usually performed by maximum entropy techniques \cite{Jarrell1996}. We use a standard maximum entropy implementation \cite{bryan} to calculate the interacting density of states for real frequencies.

\begin{figure}
\includegraphics[width=\columnwidth]{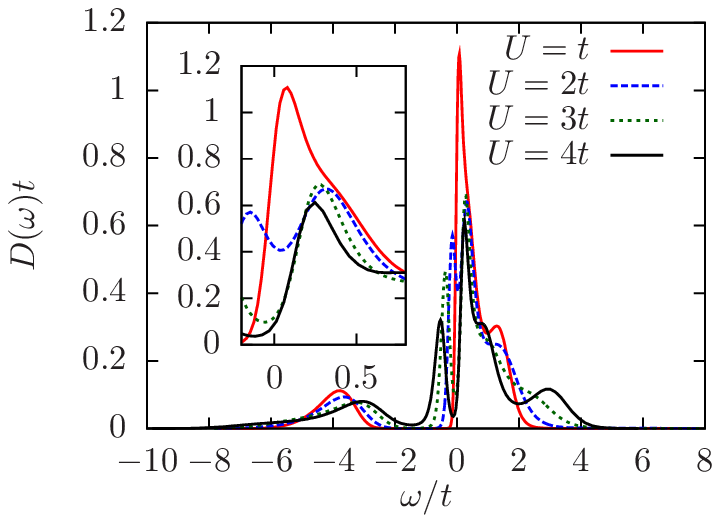}
\includegraphics[width=\columnwidth]{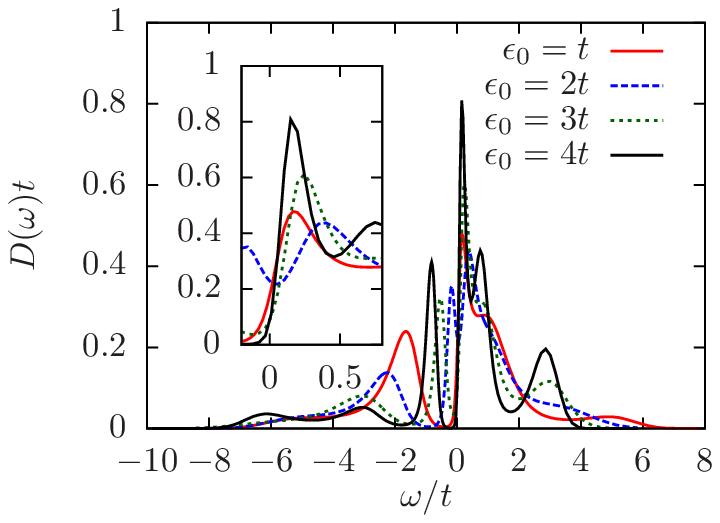}
\caption[fig]{(Color online). Density of states $D(\omega)$ at the site next to the impurity at $\beta t=10$ calculated by DMFT in combination with QMC. The analytic continuation was performed by maximum entropy. Top: density of states for $\epsilon_0/t = 3$ and different values of $U/t$. Bottom: density of states for $U/t = 4$ and different values of $\epsilon_0 / t$. The insets show a blowup of the region near the Dirac energy.}
 \label{dos}
\end{figure}
{\em Results.}
The density of states at the site next to the impurity is shown in Fig.~\ref{dos}. We find a resonance whose width decreases either by increasing $\epsilon_0 / t$ or by increasing $U/t$. The reduction in width by the interactions is a characteristic feature of magnetic impurity problems \cite{H93}. It indicates the decoupling of the impurity degrees of freedom from the conduction band. For $\epsilon_0\geq2t$ and $U\geq2t$ we find a second resonance to the left of the Dirac point which is shifted further to the left for increasing values of $\epsilon_0/t$ or $U/t$. There appear also satellite peaks at high energies. These peaks are related to transitions involving configurations where the resonance hosts zero or two electrons. The positions of these peaks are shifted by an amount proportional to $\epsilon_0$, confirming that they are related to the impurity.

\begin{figure}
\includegraphics[width=\columnwidth]{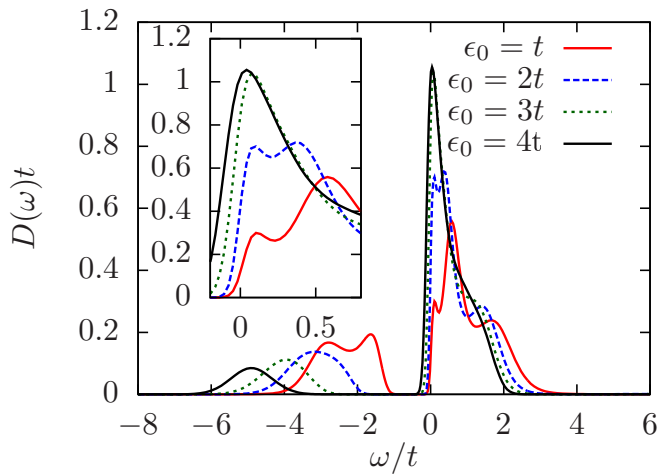}
\caption[fig]{(Color online). Density of states at the site next to the impurity at $\beta t=10$ calculated by second-order perturbation theory. Parameters are $\beta t= 10$, $U/t = 1$, and different values of $\epsilon_0 / t$. The inset is a blowup of the region near the Dirac energy.}
 \label{dosPert}
\end{figure}
We have repeated the previous calculations using an input self-energy obtained from second-order perturbation theory. The results are shown in Fig.~\ref{dosPert} and are consistent with those shown in Fig.~\ref{dos} for $U/t \lesssim 3$. One clearly sees the formation of a resonance that becomes sharper and shifts towards $\omega/t=0$ for increasing $\epsilon_0/t$. We also have a peak to the left of $\omega/t=0$ that is shifted with increasing $\epsilon_0$.

\begin{figure}
\includegraphics[width=\columnwidth]{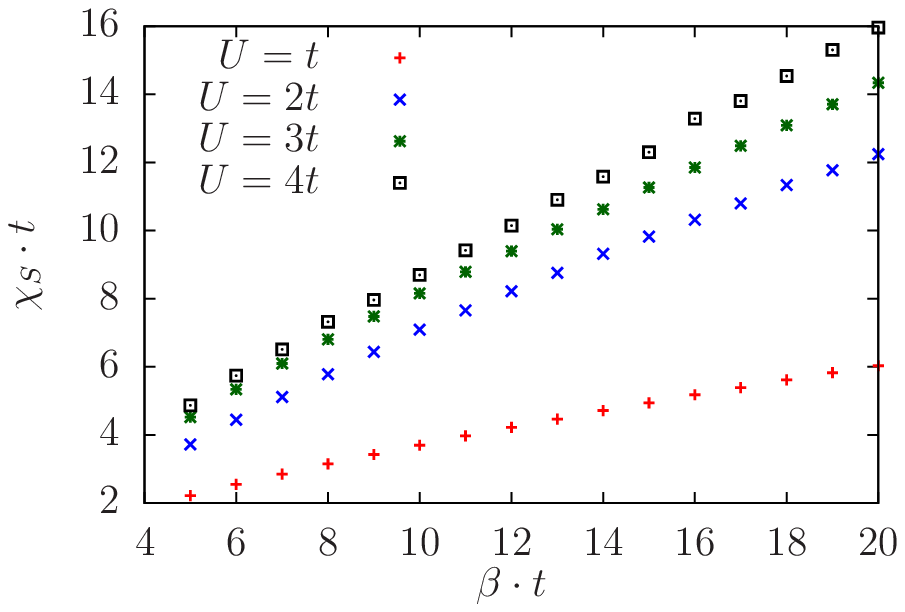}
\includegraphics[width=\columnwidth]{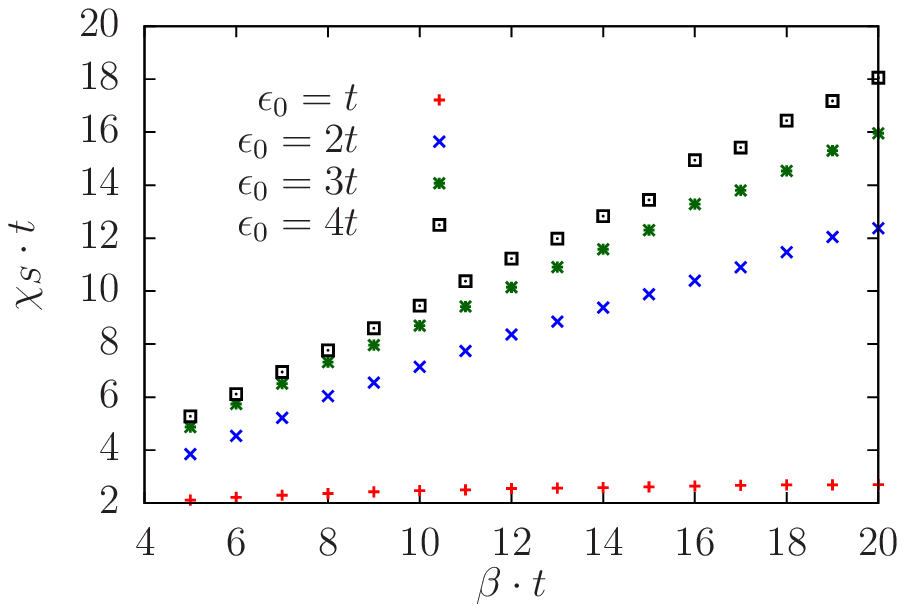}
\caption[fig]{(Color online). Local magnetic susceptibility $\chi_S$ as function of the inverse temperature calculated by DMFT in combination with QMC. Top: susceptibilities for $\epsilon_0/t = 3$ and different values of $U/t$. Bottom: susceptibilities for $U/t = 4$ and different values of $\epsilon_0 / t$.}
 \label{susc}
\end{figure}
We calculate the magnetic susceptibility of the impurity in the imaginary-time framework of the QMC via
\begin{equation}
	\chi_S = \int_0^\beta d\tau \langle(n_{\uparrow}(\tau)-n_{\downarrow}(\tau))(n_{\uparrow}(0)-n_{\downarrow}(0))\rangle \;\;,
\label{eq:spinSusc}
\end{equation}
where $\tau$ denotes imaginary time, $n_\sigma(\tau)=e^{-\tau H}n_\sigma e^{\tau H}$, and $\beta=1/k_BT$. As usual, $T$ denotes temperature and $k_B$ Boltzmann's constant. Fig.~\ref{susc} shows $\chi_S$ as function of the inverse temperature. A fully localized spin possesses the susceptibility $\chi_{loc} = \beta t$. The numerical results are consistent with a Curie dependence on temperature, $\chi \propto 1 / T$, suggesting the formation of a local moment. This moment is not fully localized at the positions closest to the impurity, and the measured $\chi_S$ is smaller than $\approx\chi_{loc} / 2$. As the interaction increases the magnetic moment becomes better defined and more localized near the impurity. We find no sign of saturation of the susceptibility down to the lowest temperatures. This result is consistent with a ferromagnetic Kondo coupling.

{\em Discussion.}
We have studied the effects of interactions in the presence of a resonance in the graphene electronic spectrum. We use a local approach, and fully include quantum fluctuations. The interaction is described by a local Hubbard term, and we do not consider the imperfect screening in graphene near the neutrality point which is expected in suspended layers. Given this interaction, our calculation can be regarded as the solution of a quantum cluster problem. The effect of interactions in the medium surrounding to the cluster is described by means of a local self-energy. The results obtained when this self-energy is calculated by DMFT and by perturbation theory are mutually consistent, in the regime where perturbation theory is valid. The off diagonal corrections to the self-energy not included here are mostly due to the long range part of the interaction \cite{GGV94}. They lead to an increase in the Fermi velocity\cite{Eetal11}. The density of states at low energies is reduced, favoring further the formation of local moments.

 We find that interactions reduce the width of the resonance induced by the impurity, and lead to a magnetic susceptibility which grows at low temperatures as $\chi_S \propto 1/T$. The local moment is not quenched at the lowest temperatures studied. This is consistent with the existence of a ferromagnetic coupling between the moment and the valence electrons. Note that the exchange mechanism which leads to an antiferromagnetic interaction for magnetic impurities coupled to a conduction band does not exist in the case of a resonance built up from the orbitals which also give rise to the conduction band. The existence of the vacancy does not lead to virtual hoppings between extended and localized states. An electron occupying the resonant state interacts with a conduction electron through the onsite Hubbard term. This coupling favors a ferromagnetic alignment of the spin of the electron in the resonance and the spin of the conduction electrons.

 The scale at which the local moment studied here leads to significant effects depends on the value of $U/t$, which is not very precisely determined in graphene. Calculations based on the Local Density Functional Approximation suggest\cite{PCMH07} $U/t \sim 1$, while quantum many body calculations for aromatic molecules give \cite{PCR50,VSCL10,Wetal11} $U/t \sim 3$. The value of $U/t$ is bounded by $U/t \approx 4.5$, above which graphene should become antiferromagnetic \cite{ST92}.

{\em Acknowledgements.}
We appreciate helpful discussions with M. I. Katsnelson and A. H. Castro Neto. F. G. and H. O. are supported by  by MICINN (Spain), grants FIS2008-00124 and CONSOLIDER CSD2007-00010.
P.H., S.F. and T.P. acknowledge financial support by the Deutsche Forschungsgemeinschaft through the collaborative research center SFB 602.
This work was supported by the German Academic Exchange Service (DAAD).
We used computational resources provided by the Gesellschaft f\"ur wissenschaftliche Datenverarbeitung G\"ottingen (GWDG).
Our implementation of the DMFT, QMC, and maximum entropy algorithms is based in the libraries \cite{alps} and DMFT application \cite{alpsdmft} of the ALPS project. ALPS (Applications and Libraries for Physics Simulations) is an open source effort providing libraries and simulation codes for strongly correlated quantum mechanical systems.

\bibliography{vacancy,literature}
\end{document}